\documentclass{iau_FM}
\usepackage{graphicx}

\title[Amplitude Variability in $\gamma$ Dor and $\delta$ Sct Stars]{Amplitude Variability in $\gamma$ Dor and $\delta$ Sct Stars Observed by the {\it Kepler} Spacecraft}

\author[Guzik, Kosak,  Bradley \& Jackiewicz]   %% give here short author list %%
{Joyce A. Guzik$^1$, Katie Kosak$^{1,2}$, Paul A. Bradley$^1$, 
 \and Jason Jackiewicz$^3$}

\affiliation{$^1$Los Alamos National Laboratory, Los Alamos, NM USA 87545 \\ email: {\tt joy@lanl.gov} \\[\affilskip]
$^2$Florida Institute of Technology, Melbourne, FL USA 32901 \\[\affilskip]$^3$New Mexico State University, Las Cruces, NM USA 88003}

\pubyear{2015}
\jname{Astronomy in Focus, Volume 2} 
\editors{Piero Benvenuti, ed.}
\begin{document}

\maketitle

\begin{abstract}

The NASA {\it Kepler }spacecraft data revealed a large number of multimode nonradially pulsating $\gamma$ Dor and $\delta$ Sct variable star candidates.
The {\it Kepler} high precision long time-series photometry makes it possible to study amplitude variations of the frequencies. We summarize recent literature on amplitude and frequency variations in pulsating variables. We are searching for amplitude variability in several dozen faint  $\gamma$ Doradus or $\delta$ Scuti variable-star candidates observed as part of the {\it Kepler} Guest Observer program.  We apply several methods, including a Matlab-script wavelet analysis developed by J. Jackiewicz, and the wavelet technique of the VSTAR software (http://www.aavso.org/vstar-overview).  Here we show results for two stars,  KIC 2167444 and KIC 2301163.  We discuss the magnitude and timescale of the amplitude variations, and the presence or absence of correlations between amplitude variations for different frequencies of a given star.  Amplitude variations may be detectable using {\it Kepler} data even for stars with {\it Kepler} magnitude $>$14 with low-amplitude frequencies ($\sim$100 ppm) using only one or a few quarters of long-cadence data. We discuss proposed causes of amplitude spectrum variability that will require further investigation.

\end{abstract}

\firstsection % if your document starts with a section,
              
\section{Why are amplitude variation unexpected and important?}

For single stars with pulsations unstable to a driving mechanism such as the $\kappa$ mechanism, pulsation properties are determined by the structure of the star, which usually changes very slowly with time via evolutionary processes, e.g., nucleosynthesis for main-sequence stars, or cooling for white dwarfs.  The timescales for these processes are hundreds to thousands of years, rather than the hours to years over which we have photometric data and detect significant variations.  Since growth rates calculated by linear nonadiabatic pulsation codes can be large (normalized work 10$^{-3}$--10$^{-6}$ per period), pulsation amplitudes should grow relatively quickly to reach a limiting amplitude.  On the other hand, stochastically excited pulsations, as found in solar-like and red giant stars, are continuously excited and damped, so their amplitudes are expected to vary.

Amplitude variations may be useful diagnostics of energy partition/exchange between modes, and may involve modes of degree $l$ $\ge$ 3 that are more difficult to observe in photometry, or gravity modes with high amplitude in the stellar interior that aren't visible at the surface, or even stable modes.  These variations may tell us something about energy exchange with internal dynamical processes (convection, rotation, magnetic fields) or changes in ionization regions that we cannot observe directly.  They may indicate interaction with the external environment, e.g., via mass outflow, accretion, or tidal forces from a binary companion or planet.  Understanding the mechanisms that limit amplitudes or cause amplitude variations will be important to understand mode selection, and to validate nonlinear nonradial and nonadiabatic pulsation models.

\begin{figure}[b]
% \vspace*{-2.0 cm}
\begin{center}
 \includegraphics[width=3.4in]{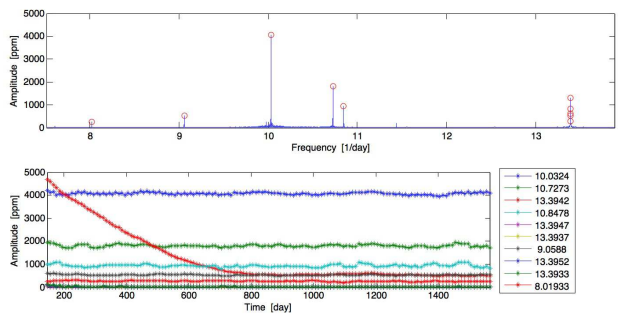} 
% \vspace*{-1.0 cm}
 \caption{Matlab-script generated KIC 7106205 amplitude spectrum (top) and amplitude vs. time for ten highest-amplitude modes.  The mode at 13.3942 c/d shows a dramatic amplitude decrease during the first $\sim$ 600 days, as first reported by \cite[Bowman \& Kurtz (2014)]{bowmankurtz2014}.}
  \label{7106205}
\end{center}
\end{figure}

 \begin{figure}[b]
% \vspace*{-2.0 cm}
\begin{center}
 \includegraphics[width=4.0in]{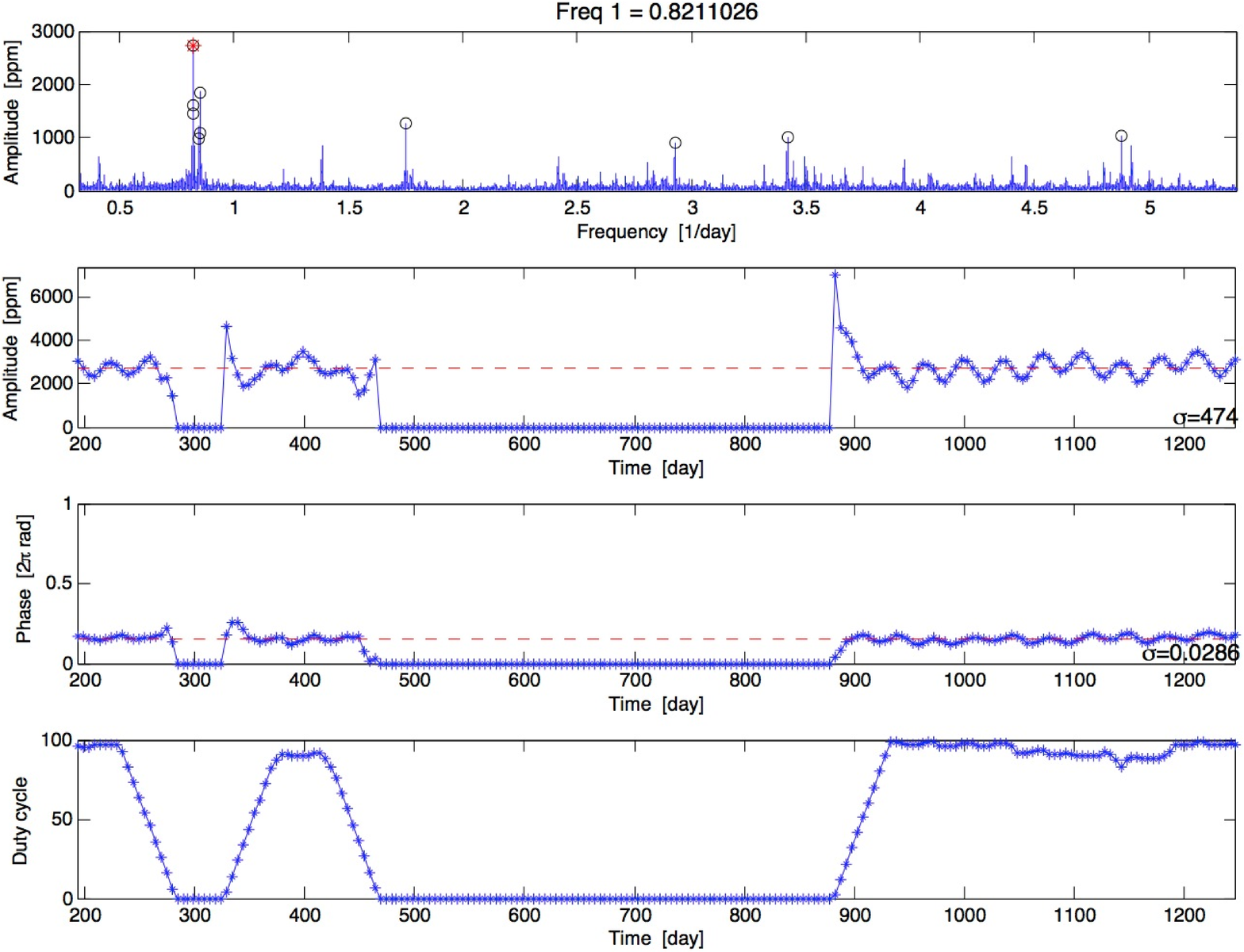}
% \vspace*{-1.0 cm}
 \caption{Matlab-script amplitude spectrum (top panel) and wavelet analysis for amplitude (second panel) and phase (third panel) variations, and duty cycle (bottom panel) for highest amplitude 0.8211 c/d mode for KIC 2167444.  Although we have applied a correction for decreasing duty cycle, the amplitudes are still unreliable very near the time regions where the duty cycle decreases to zero.  The amplitude and phase variations show the same periodicity but are offset in phase by $\sim$90$^{\circ}$, as is expected for beating between two modes according to \cite[Breger \& Pamyatnykh (2006)]{breger2006}. }
   \label{Matlab2167444}
\end{center}
\end{figure}

 \begin{figure}[b]
% \vspace*{-2.0 cm}
\begin{center}
 \includegraphics[width=3.4in]{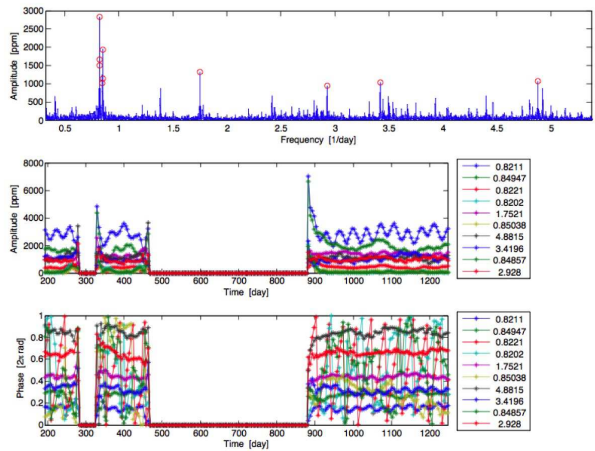} 
% \vspace*{-1.0 cm}
 \caption{Matlab-script amplitude spectrum (top panel) and wavelet analysis for amplitude variation (middle panel) and phase variation (bottom panel) of 10 highest-amplitude modes of KIC 2167444.  The four modes with rapid phase variation are spaced about 0.001 c/d from the main peaks at 0.8211, 0.8495 c/d main peak, and are likely artifacts of the evenly spaced 30-minute cadence light curve that is oversampled by a factor of 10.}
   \label{Matlab2167444allmodes}
\end{center}
\end{figure}

 \begin{figure}[b]
% \vspace*{-2.0 cm}
\begin{center}
 \includegraphics[width=3.4in]{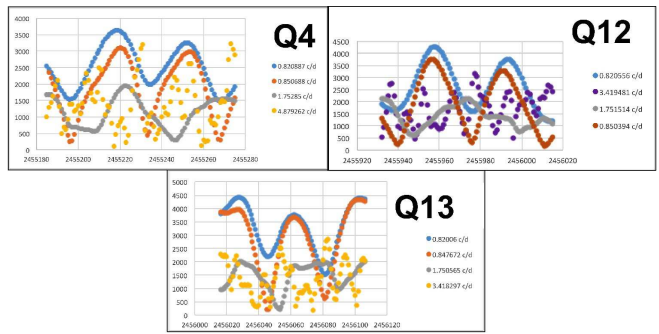} 
% \vspace*{-1.0 cm}
 \caption{VSTAR wavelet analysis for KIC 2167444 Q4, Q12, and Q13 data.  The highest-amplitude modes at 0.821 (blue dots) and 0.851 (orange dots) c/d show $\sim$35-day amplitude modulations that are not exactly in phase with each other.  The next two highest-amplitude modes show quasi-periodic amplitude variations.}
   \label{VSTAR2167444}
\end{center}
\end{figure}

 \begin{figure}[b]
% \vspace*{-2.0 cm}
\begin{center}
 \includegraphics[width=3.4in]{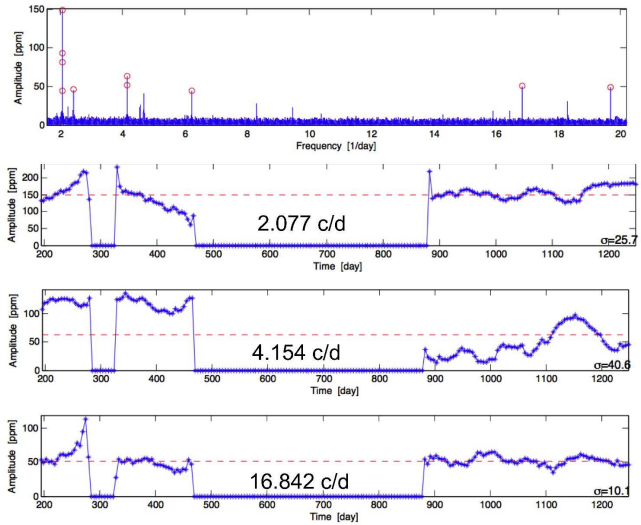} 
% \vspace*{-1.0 cm}
 \caption{Matlab-script amplitude spectrum (top panel) and wavelet analysis (next three panels) for highest-amplitude modes of KIC 2301163.  The 2.077 c/d mode and its possible harmonic at 4.154 c/d do not show the same amplitude variation.  The $\delta$ Sct-type p mode at 16.842 c/d does not show significant amplitude variation.}
   \label{KIC2301163modes}
\end{center}
\end{figure}

\section{Causes of amplitude/frequency variations}

Amplitude and/or frequency variations have been found among nearly all types of non-stochastically excited pulsating variables: $\delta$ Sct (\cite{breger2012, breger2006, bowmankurtz2014}); $\gamma$ Dor (\cite{rostopchina2013}); $\beta$ Cep (\cite{pigulski2008}), roAp (\cite{balona2013, medupe2015}); classical pulsators such as high-amplitude $\delta$ Sct stars (\cite{zhou2011, khokhuntod2011}), Cepheids (\cite{engle2015}), RR Lyrae (\cite{chadidpreston2013}), Miras and yellow supergiants (\cite{percyyook2014, percykhatu2014}); white dwarfs (DBV; \cite{handler2003} and DAV; \cite{bell2015}); GW Vir stars (\cite{vauclair2012}), sdB stars (\cite{kilkenny2010, langfellner2012}), and an extreme helium subdwarf (\cite{bearsoker2014}).

A non-comprehensive list of proposed explanations for the variations includes:  parametric instability (unstable high frequency mode $\nu_1$ excites two lower frequency stable modes with $\nu_2$ + $\nu_3$ = $\nu_1$; see \cite{dziembowski1985}); resonant mode coupling (\cite{forteza2015}); stochastic excitation instead of intrinsically unstable modes (see, e.g., \cite{huber2011}); energy from the modulated mode exchanged with, e.g., the convection zone, ionization region or magnetic field; `weather' (atmospheric disturbances) from a tidally locked planet (\cite{bearsoker2014}); tidal effects from an unseen binary or planetary companion; outbursts (e.g., Be star 102719279; \cite{soto2010}), or accretion (e.g., GW Lib; \cite{toloza2015}) changing the star's structure; pulsations sampling the crystallization region (white dwarf interior; \cite{hermes2015}); diffusive settling of helium (\cite{cox1998}); the star is caught during a phase of rapid evolution, e.g., at the edge of an instability strip, or during the rapid core contraction phase at the end of core hydrogen burning.  Sometimes apparent amplitude variations are attributable to insufficiencies in the time series data or analysis, e.g., to very close frequencies not being resolved, an interruption in the time series, or an artifact from the temporal distribution of data.  The papers by \cite[Bowman \& Kurtz (2014), Breger \& Montgomery (2014), Barcel{\'o} Forteza \etal\ (2015), Holdsworth \etal\ (2014)]{bowmankurtz2014, bregermontgomery2014, forteza2015, holdsworth2014}, and \cite[Percy \& Khatu (2014)]{percykhatu2014} provide discussion and further references for causes of amplitude and frequency variations.

\section{Amplitude variations in $\gamma$ Dor/$\delta$ Sct stars}

Detecting (or ruling out) amplitude variations requires high-precision continuous time series data that captures many pulsation periods.  We now have many data sets that can be used to study such variations, including those from {\it Kepler} and CoRoT ($\delta$ Sct and $\gamma$ Dor stars), ASAS (B stars), WET (white dwarfs and roAp stars), and AAVSO (long-period variables and giants) observations.  The {\it Kepler} spacecraft has returned continuous time-series observations  spanning months to years, either in long cadence (30-minute integrations per data point), or short cadence (1-minute integrations), with micro-magnitude precision in the amplitude spectrum.

We are searching for amplitude variations in mostly faint (K$_p$ mag $>$14) $\gamma$ Dor and $\delta$ Sct candidates discovered in long-cadence data via the {\it Kepler} Guest Observer program.  We used the weighted-wavelet z-transform technique (Foster 1996)  available in the VSTAR software from the American Association of Variable Star Observers.  For the example analyses shown below, we use $\sim$1000 data points per Gaussian wavelet, that translate to $\sim$ 20-day windows for long-cadence data.  We also show results for the wavelet-analysis script written using Matlab by J. Jackiewicz, choosing 50-day wide data windows, with 10-day or 5-day offsets.  We note that it is important to consider the width of the time series to analyze, as well the offset between windows; too-large windows and too-large offsets will average out or decrease the size of amplitude variations, and too-small windows will have low signal-to-noise and fail to resolve closely spaced modes.

We first tested the Matlab script on {\it Kepler} data for KIC 7106205 (K$_p$ mag = 11.455), and easily detect in long-cadence data the large amplitude decrease of the 13.3942 c/d mode reported by \cite[Bowman \& Kurtz (2014)]{bowmankurtz2014} (Fig. \ref{7106205}).  This script corrects the amplitude and phase vs. time for reduced duty cycle in the {\it Kepler} data.  As seen in later analysis, this correction is only partially successful at the edges of observation quarters without data (zero duty cycle).
\subsection{KIC 2167444}

KIC 2167444 is a K$_p$ mag 14.1 $\gamma$ Dor candidate observed by {\it Kepler} in long cadence during Quarters 2, 4, and 10-13.  Figure \ref{Matlab2167444} shows the Matlab-script amplitude spectrum (top panel) and the wavelet-analysis amplitude (2nd panel) and phase (3rd panel) vs. time for the highest-amplitude peak at 0.8211 c/d.  The amplitude and phase show a pronounced modulation of about 32 days.  The amplitude is not reliable in the regions where the duty cycle drops off quickly (bottom panel).  Figure \ref{Matlab2167444allmodes} shows the Matlab-script results for the amplitude and phase variations of the 10 highest-amplitude modes (marked with red circles on the amplitude spectrum in the top panel).  The second-highest amplitude mode at 0.8495 c/d shows an amplitude modulation similar to the 0.8211 c/d  mode during Q2 and 4, but a more gradual variation in Q10-13.  These two close frequencies should be resolved with the 50-day wavelet window, (resolution 0.02 c/d); their beating period is (1/(0.8211025-0.8494719)) = $\sim$ 35 days, so it is possible that beating between these modes is causing the amplitude variation, or that these two modes are interacting.  There is a phase shift of $\sim$90$^{\circ}$ between the amplitude and phase variation of this mode, which fulfills the criteria for two modes beating against each other, according to \cite[Breger \& Pamyatnykh (2006)]{breger2006}.  The modes at 1.7521, 4.8815, 3.4196, and 2.9280 c/d also show more subdued amplitude variations with similar periodicity; these modes are not obvious combination frequencies of the two highest-amplitude modes, so the periodicity is puzzling.  An alternative explanation is that several high-amplitude modes are coupling to a low-frequency, possibly stable mode, as suggested by \cite[Barcel\'o Forteza \etal\ (2015)]{forteza2015} to explain amplitude modulations in the $\delta$ Sct star KIC 5892969.

The pair of modes at 0.8221 and 0.8202 c/d, and the pair at 0.8486 and 0.8504 c/d show very large phase variations in Fig. \ref{Matlab2167444allmodes} (bottom panel).  These modes are all multiples of $\sim$0.001 c/d away from a main mode, and are likely side-lobe artifacts generated by the finite-length data set of 1050 days (1/1050 days = $\sim$0.001 c/d).  These modes essentially disappear when the main mode is prewhitened.

We also examined the data using the VSTAR wavelet analysis, and show results for the four highest-amplitude modes in Quarters 4, 10, and 13 (Fig. \ref{VSTAR2167444}).  VSTAR finds the same highest-amplitude modes, including the 0.82 and 0.85 c/d modes.  The VSTAR plots show that the amplitude variations between the 0.82 and 0.85 c/d highest amplitude modes are not exactly in phase with each other.  The amplitude variation of the 1.75 c/d mode is more regular during Q4 than in subsequent quarters.   In Q12 and 13, the 3.42 c/d mode has a higher amplitude, while in Q4 the 4.88 c/d mode is more prominent.  
 
 \subsection{KIC 2301163}
 
 KIC 2301163 is a K$_p$ mag 14.1 $\gamma$ Dor/$\delta$ Sct hybrid candidate that was also observed in Quarters 2, 4, and 10-13.  This star shows significant amplitude variations, even though the amplitudes are an order of magnitude lower than for KIC 2167444 ($\sim$100 ppm instead of $\sim$1000 ppm).  Figure \ref{KIC2301163modes} shows the amplitude spectrum and the amplitude vs. time for three of the highest-amplitude peaks.  The highest-amplitude mode at 2.077 c/d and the next-highest amplitude mode at 4.154 c/d, nearly exactly twice the frequency of the first mode, show different amplitude variations, even though one mode could be a harmonic of the other.  The possible $\delta$ Sct-type p mode at 16.842 c/d does not show significant amplitude variation (standard deviation 10 ppm, about the noise level of the amplitude spectrum).  

\section{Conclusions}

Amplitude variations may be detectable by {\it Kepler} even for stars with K$_p$$>$14 with low-amplitude frequencies $\sim$100 ppm using only one or a few quarters of long-cadence data. Amplitude variations for different frequencies are sometimes correlated.  It seems that analysis of the {\it Kepler} data to interpret amplitude variations requires significant effort for each star, and that it will be difficult to find patterns or draw a general conclusion by studying an ensemble of stars.

\acknowledgments

We gratefully acknowledge the NASA {\it Kepler} Guest Observer Program, the DOE Science Undergraduate Laboratory Internship program, and Los Alamos National Laboratory.

\end{document}